# Generative artificial intelligence usage by researchers at work: Effects of gender, career stage, type of workplace, and perceived barriers

Pablo Dorta-González [1,*], Alexis Jorge López-Puig [2,3], María Isabel Dorta-González [4], Sara M. González-Betancor [5]

[1] Institute of Tourism and Sustainable Economic Development (TIDES), Campus de Tafira, University of Las Palmas de Gran Canaria, 35017 Las Palmas de Gran Canaria, Spain. ORCID: 0000-0003-0494-2903

[2] Agencia Canaria de Calidad Universitaria y Evaluación Educativa (ACCUEE), C/ Álamo, 54- 2º Planta, Gobierno de Canarias, 35014 Las Palmas de Gran Canaria, Spain. ORCID: 0000-0002-8254-3513

[3] Departament of Mathematics, Campus de Tafira, University of Las Palmas de Gran Canaria, 35017 Las Palmas de Gran Canaria, Spain.

[4] Department of Computer and Systems Engineering, Avenida Astrofísico Francisco Sánchez s/n, University of La Laguna, 38271 La Laguna, Spain. ORCID: 0000-0002-7217-9121

[5] Department of Quantitative Methods in Economics and Management, Campus de Tafira, University of Las Palmas de Gran Canaria, 35017 Las Palmas de Gran Canaria, Spain. ORCID: 0000-0002-2209-1922

[*] Correspondence: pablo.dorta@ulpgc.es

## Highlights

- Usage of generative AI tools is lower among women (7%), advanced researchers (19%), and individuals without specific training (8%).
- Encountering barriers is associated with an 11% increase in generative AI tool usage.
- Higher usage is observed among researchers in for-profit companies (19%), medical research (16%), and hospitals (15%).
- Government advisors utilize generative AI tools 45% more frequently than those in typical government roles.

## Abstract

The integration of generative artificial intelligence technology into research environments has become increasingly common in recent years, representing a significant shift in the way researchers approach their work. This paper seeks to explore the factors underlying the frequency of use of generative AI amongst researchers in their professional environments. As survey data may be influenced by a bias towards scientists interested in AI, potentially skewing the results towards the perspectives of these researchers, this study uses a regression model to isolate the impact of specific factors such as gender, career stage, type of workplace, and perceived barriers to using AI technology on the frequency of use of generative AI. It also controls for other relevant variables such as direct involvement in AI research or development, collaboration with AI companies, geographic location, and scientific discipline. Our results show that researchers who face barriers to AI adoption experience an 11% increase in tool use, while those who cite insufficient training resources experience an 8% decrease. Female researchers experience a 7% decrease in AI tool usage compared to men, while advanced career researchers experience a significant 19% decrease. Researchers associated with government advisory groups are 45% more likely to use AI tools frequently than those in government roles. Researchers in for-profit companies show an increase of 19%, while those in medical research institutions and hospitals show an increase of 16% and 15%, respectively. This paper contributes to a deeper understanding of the mechanisms driving the use of generative AI tools amongst researchers, with valuable implications for both academia and industry.

## 1. Introduction

The Nature survey on Artificial Intelligence (AI) and Science was conducted in September 2023, and the results were published in October of that year (Van Noorden and Perkel, 2023). The survey found that scientists were both concerned and excited about the increasing use of artificial intelligence tools in research. More than half of the respondents expected these tools to be very important or essential to their fields in the next decade. However, they also expressed strong concerns about how artificial intelligence is changing the way science is conducted, such as the risk of recognizing patterns without understanding them, bias or discrimination in the data, fraud, or irreproducibility (Van Noorden and Perkel, 2023). The survey also found that the use of AI tools varied by discipline, with computer science, engineering, and mathematics being the most frequent users, followed by physics, chemistry, and biology. The least frequent users were in social sciences, humanities, and arts. The survey has not been repeated since, but Nature has continued to publish articles on the impact of artificial intelligence on science (Nature, 2023c).

Despite the apparent benefits, concerns are emerging about the ethical and methodological implications of integrating AI into research (Bin-Nashwan et al., 2023; ERC, 2023; Nature, 2023b). Concerns focus on the potential proliferation of unreliable research, bias in data, increased risk of fraud, and challenges to traditional scientific standards. To address these concerns, calls have been made for transparency, ethical guidelines, and enhanced oversight mechanisms to ensure the responsible use of AI while maintaining the integrity and credibility of research. By adopting these strategies, researchers seek to harness the enormous potential of AI while effectively managing its associated risks and ensuring its ethical and responsible integration into scientific practices (Van Noorden and Perkel, 2023; ERC, 2023).

However, the survey results may be biased because scientists with a pre-existing interest in AI were more likely to participate. This could bias the results by overemphasizing the perspectives of AI researchers. To mitigate this problem, the current study uses a robust methodology that goes beyond descriptive data. It uses an inference technique -known as regression analysis- to isolate effects. This method allows researchers to tease apart the independent influence of specific factors, such as gender, career stage, the type of scientific institution at which a researcher works, and even scientists' perceived barriers to using this technology, on the frequency of use of generative AI. In addition, the study controls for other potentially confounding variables, including direct involvement in AI research or development, collaboration with AI-focused companies, geographic location, and scientific discipline. This holistic approach provides a more complete understanding of the factors shaping the adoption of generative AI within the broader scientific community. By considering a wide range of influences, it provides insights that go beyond the perspectives commonly found within AI enthusiast circles, thereby enriching our understanding of this phenomenon.

## 2. Literature Review

### *2.1 Artificial Intelligence and Scientific Research*

Artificial Intelligence (AI) is at the forefront of transformative technologies, revolutionizing various areas of technology and society (Briganti and Le Moine, 2020; Wang et al., 2023). Its overarching goal is to emulate human intelligence, enabling the execution of tasks at an accelerated rate compared to human capabilities (Xu et al., 2021). This capability addresses challenges such as labour shortages and mitigates individual exposure to hazardous environments (Gao et al., 2021).

A subset of AI is natural language processing (NLP), which aims to replicate human conversation and thereby enhance machine-human communication (Holler and Levinson, 2019; Trenfield et al., 2022). This allows for improved access to machines and digital content, thus democratizing access to technology. Traditionally, interacting with machines has required programming skills, creating a barrier for researchers interested in using AI to address scientific challenges. However, the development of large language models (LLMs) represents a breakthrough in NLP, making it accessible to a broader audience (De Angelis et al., 2023; Agathokleous et al., 2024). This accelerated retrieval process has the potential to accelerate discovery and development in the 21st century.

### *2.2 Large Language Models and Scientific Writing*

The advent of AI offers promising ways to save time, and the launch of ChatGPT by OpenAI in November 2022, marks a significant milestone in the development of AI language models, a field that has been under development for years (Biswas, 2023). These models, rooted in generative pretrained transformer (GPT) technology and natural language processing (NLP), enable seamless communication between computers and humans by understanding and generating human language (Hutson, 2022). Using extensive text data and neural network programming, ChatGPT and other large language models (LLMs) predict appropriate text responses based on input, facilitating human interaction (Huang and Tan, 2023). LLMs use neural networks trained on vast amounts of textual and visual data, searching for connections to accumulate knowledge that includes patterns, facts, and grammar rules (Thirunavukarasu et al., 2023). This enables autonomous text generation, query response, and task assistance, much like accessing a vast digital repository (Floridi and Chiriatti, 2020).

LLMs show promise in generating new content, particularly in the medical domain, and in supporting the automation of writing tasks. For example, ChatGPT has been shown to support clinical decision-making (Kung et al., 2023). In academia, LLMs have facilitated the writing of review articles, the design of experimental procedures, and the posing of key questions in various scientific disciplines (Marquez et al., 2023; Norris, 2023; Rahimi et al., 2023; Agathokleous et al., 2024). However, to date, LLMs have not been able to autonomously produce data-driven, original research articles from conception to publication. The production of such articles requires significant resources, including expertise, equipment, and materials. If LLMs can take on this task, they have the potential to revolutionize research practices, not only streamlining information retrieval but also surpassing traditional tasks such as writing literature reviews (Grace et al., 2024; Gückman and Zhang, 2024).

The potential applications of AI are vast, as it can perform tasks in seconds that would take considerable time and effort for most human users (Grace et al., 2024; Gückman and Zhang, 2024). Recently, a significant discourse has emerged within the research community regarding the integration of AI into scientific writing. While some advocate for AI as a valuable writing aid, many scientists and publishers oppose the idea of AI solely producing papers or being credited as the sole author (Altmäe et al., 2023; Chen, 2023; Lee, 2023; Salvagno et al., 2023). Nevertheless, AI can undoubtedly assist with academic writing; LLMs such as OpenAI's ChatGPT, Google's Gemini, and Microsoft's Copilot excel at providing grammar, vocabulary, and writing style assistance. In addition, AI resources can perform plagiarism checks and serve as literature search engines, critical tools for researchers preparing manuscripts (Huang and Tan, 2023). This potential time-saving aspect is particularly beneficial for non-native authors.

However, the use of AI in scientific writing requires scrutiny and caution. Cases of AI misuse, such as the generation of fictitious court citations leading to legal consequences, underscore the importance of responsible AI use (Neumeister, 2023). In addition, AI has limitations, such as potential copyright infringement and the generation of "artificial hallucinations" (Salvagno et al., 2023), the production of biased or inaccurate results, and the inability to discern the meaning of different sources (Alkaissi and McFarlane, 2023). Notably, AI has already deceived human reviewers by producing believable abstracts, raising concerns about its ability to produce publishable full-length scientific reviews (Gao et al., 2023).

## 3. Methodology

### *3.1 Data*

We use survey data from the Nature survey on "AI and science: what 1,600 researchers think" to investigate the frequency of use of generative AI amongst researchers at work. The raw data are available as open data for further exploration. Specifically, the raw data are provided as open supplementary information in the publication by Van Noorden and Perkel (2023). This dataset provides valuable insights into researchers' attitudes, perceptions, and usage patterns of AI technologies in the scientific domain. The survey covers a wide range of disciplines and geographic regions, providing a comprehensive representation of the research community.

Nature journalists emailed more than 40,000 scientists who had published in the last four months of 2022 and invited Nature Briefing readers to participate in the survey. The response rate for the 40,000 scientists was approximately 5%, but not all respondents completed the survey: a total of 2,728 responses were received, of which 1,801 were complete. After further removing data from inactive researchers and those who didn't specify whether they were studying, using, or not using AI, 1,659 active researchers remained. These participants were then grouped into regional categories based on where they live: Asia, North America, Europe, and Other.

### *3.2 Methods*

The survey results may be biased because scientists interested in AI were more likely to participate, potentially skewing the results toward the views of AI researchers. To address this, the study uses a robust methodology. It isolates the impact of specific factors, such as gender, career stage, job type, and scientists' perceived barriers to using this technology, on the frequency of generative AI use. It also controls for other relevant variables, such as direct involvement in AI research or development, collaboration with AI companies, geographic location, and scientific discipline. This comprehensive approach allows researchers to gain a clearer understanding of how different factors influence the use of generative AI amongst scientists.

The Chi2 test is a nonparametric test, which means that makes no assumptions about the distribution of the variables. The test is based on the chi-squared statistic, which is calculated by comparing the observed frequencies of the categories of variables with the expected frequencies. The expected frequencies are calculated under the assumption that there is no association between the variables. The chi-squared statistic is then used to calculate a p-value, which is the probability of obtaining a value as extreme as -or more extreme than- the observed chi-squared statistic if there is no association between the variables. A low p-value (less than the significance level) indicates that there is a statistically significant association between the variables.

In the context of the variable Q21, the Chi2 test can be used to determine if there is an association between Q21 and any of the other variables, and to identify the variables that are most strongly associated with Q21. It is important to note that the Chi2 test does not indicate the direction of the causal relationship between the variables. For example, if the Chi2 test shows that there is an association between Q21 and gender, this does not mean that Q21 causes gender or that gender causes Q21. It is possible that there is a third variable that causes both Q21 and gender. Therefore, although the Chi2 test serves as a powerful tool for evaluating the correlation between two categorical variables, it is crucial to complement its use with other statistical techniques to gain a full understanding of the relationship between the variables.

To examine the determinants of the frequency of use of generative AI amongst researchers, we employ a regression model. Specifically, we use Ordinary Least Squares (OLS) regression analysis, with robust standard errors estimated using the hc3 method, as suggested by Davidson and MacKinnon (1993, 2004), who report that this often produces better results when the model is heteroskedastic. Given the potential problems of non-normality and heteroscedasticity in the data, the use of robust standard errors helps to mitigate bias in the parameter estimates. Our regression model includes the frequency of use of generative AI as the dependent variable and employs a set of independent variables representing various factors hypothesized to influence researchers' use of generative AI.

We also log-transform the endogenous variable to reduce skewness and improve model fit. In addition, the inclusion of robust standard errors ensures the validity of statistical inference in the presence of heteroskedasticity and potential model misspecification. Overall, by employing OLS regression with robust standard errors and log-transformed endogenous variables, we aim to provide robust estimates of the factors influencing the frequency of use of generative AI amongst researchers, thus contributing to a deeper understanding of the dynamics of AI adoption within the scientific community.

## 4. Results

### *4.1 Descriptive Analysis of the Survey Data in General and the Degree of Association with the Use of Generative AI*

Figure 1 illustrates the use of AI tools amongst survey respondents. Question Q21 assessed the frequency of using tools such as ChatGPT, GPT-4, PALM, MidJourney, LLaMA, or any related products in the workplace. Responses were measured on a scale of 1 to 5, with 5 being the highest frequency of use. There was considerable variability in responses to this question amongst the 1,649 survey respondents. Specifically, of the 1,649 active research participants shown in Figure 1, the majority (550, i.e. 33%) reported never using AI tools in their work. A significant portion (448, i.e. 27%) reported using them only a few times. Occasional use was also reported by 364 respondents (22%). Conversely, a smaller group reported more frequent use: 184 (11%) used AI tools more than once a week, and 103 (6%) used them daily.

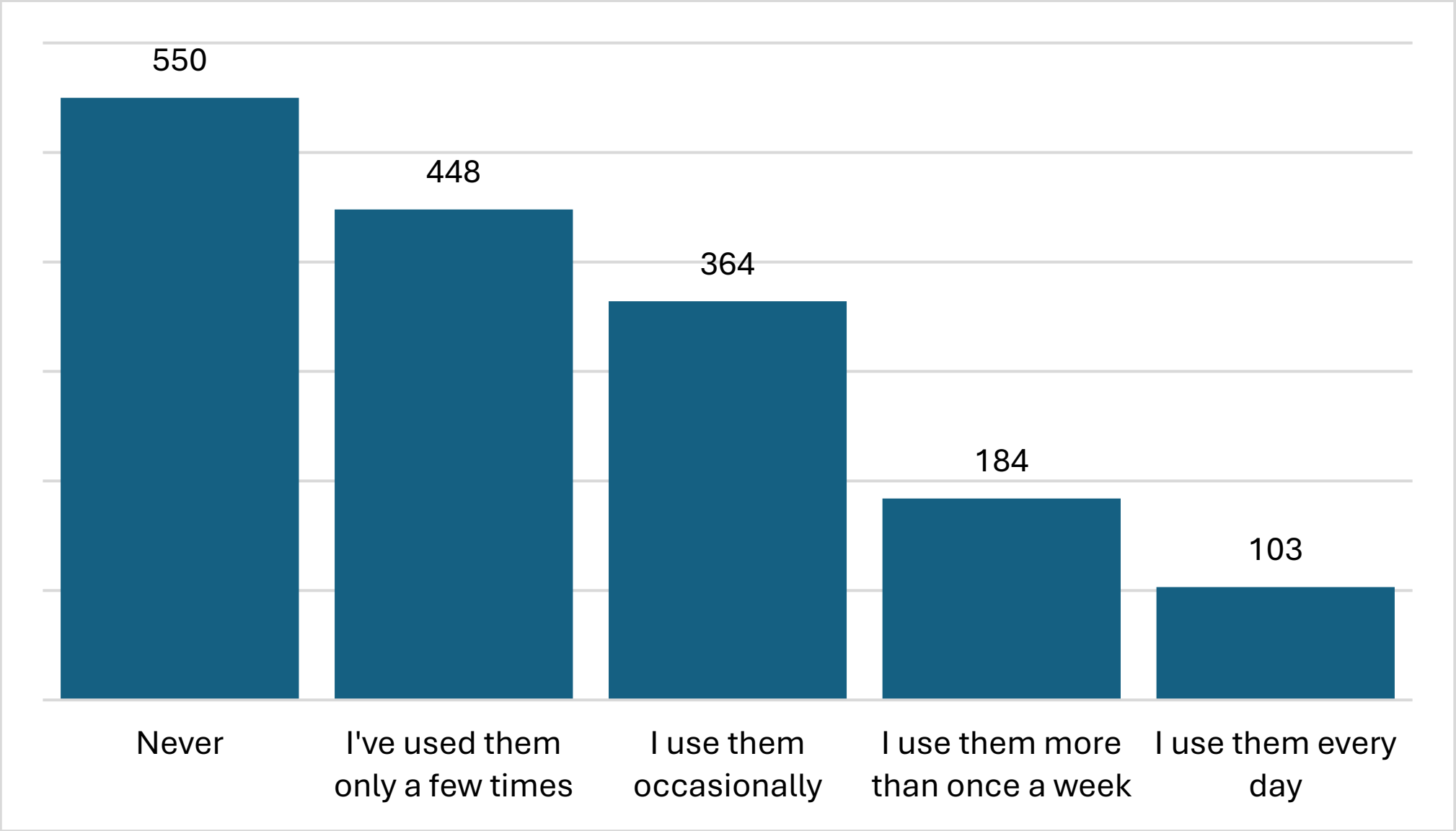

Figure 1. Distribution of the question Q21: How often do you use these tools (e.g., ChatGPT, GPT-4, PALM, MidJourney, LLaMA, or any products built using these tools) at work

To examine the relationship between this variable and other study factors, a chi-square test was conducted, which revealed a significant association between AI tool usage and other measured variables (Table 1).

The results highlight perceived barriers to AI adoption (39%), with many researchers citing a lack of skills (22%), funding (21%), computing power (19%), data availability (18%), and training resources (17%). These challenges, which span multiple variables, suggest barriers to AI integration across scientific fields.

A notable number of respondents were involved in collaborations with scientists from AI development companies (14%), underscoring the active engagement between academia and industry. In addition, the perceived value of these collaborations varied, with a significant proportion considering them somewhat or very important (17%) for advancing AI research.

The gender distribution of respondents was skewed toward males (71%), while the geographic distribution showed a diverse representation across regions, with a majority from Europe (32%), Asia (27%), and North America (19%). The career stage distribution was relatively balanced between advanced-career (39%), mid-career (31%), and early-career (25%). There was observed variation in the work organization distribution, with a substantial majority affiliated with universities (64%). Finally, the distribution of research fields showed a diverse representation across scientific disciplines.

Table 1. Descriptive analysis of the survey. Frequencies of categories and Chi2 test to determine if there is a significant relationship between Q21 and the rest of the variables.

| | Variable | Obs | Mean | Std. dev. | Min | Max | Chi2 | Prob |
|---|---|---|---|---|---|---|---|---|
| Q21: How often do you use these tools (e.g., ChatGPT, GPT-4, PALM, MidJourney, LLaMA, or any products built using these tools) at work? | | | | | | | | |
| 1 | Never | 550 | 0.334 | 0.472 | 0 | 1 | | |
| 2 | I've used them only a few times | 448 | 0.272 | 0.445 | 0 | 1 | | |
| 3 | I use them occasionally | 364 | 0.221 | 0.415 | 0 | 1 | | |
| 4 | I use them more than once a week | 184 | 0.112 | 0.315 | 0 | 1 | | |
| 5 | I use them every day | 103 | 0.062 | 0.242 | 0 | 1 | | |
| Q1: Does your research work directly involve studying or developing AI? | | | | | | | 189.69 | 0.000 |
| 0 | No | 856 | 0.519 | 0.500 | 0 | 1 | | |
| 1 | Yes | 793 | 0.481 | 0.500 | 0 | 1 | | |
| Q7: Do you feel that there are barriers preventing you, or your research team, from developing or using AI as much as you would like? | | | | | | | 285.43 | 0.000 |
| 0 | No | 630 | 0.382 | 0.486 | 0 | 1 | | |
| 1 | Yes | 644 | 0.391 | 0.488 | 0 | 1 | | |
| 2 | Missing flag | 375 | 0.227 | 0.419 | 0 | 1 | | |
| Q8a: Barriers: Lack of skills or skilled researchers | | | | | | | 270.77 | 0.000 |
| 0 | No | 907 | 0.550 | 0.498 | 0 | 1 | | |
| 1 | Yes | 367 | 0.223 | 0.416 | 0 | 1 | | |
| 2 | Missing flag | 375 | 0.227 | 0.419 | 0 | 1 | | |
| Q8b: Barriers: Lack of training resources | | | | | | | 272.74 | 0.000 |
| 0 | No | 990 | 0.600 | 0.490 | 0 | 1 | | |
| 1 | Yes | 284 | 0.172 | 0.378 | 0 | 1 | | |
| 2 | Missing flag | 375 | 0.227 | 0.419 | 0 | 1 | | |
| Q8c: Barriers: Lack of computing resources | | | | | | | 281.82 | 0.000 |
| 0 | No | 960 | 0.582 | 0.493 | 0 | 1 | | |
| 1 | Yes | 314 | 0.190 | 0.393 | 0 | 1 | | |
| 2 | Missing flag | 375 | 0.227 | 0.419 | 0 | 1 | | |
| Q8d: Barriers: Lack of funding | | | | | | | 276.38 | 0.000 |
| 0 | No | 921 | 0.559 | 0.497 | 0 | 1 | | |
| 1 | Yes | 353 | 0.214 | 0.410 | 0 | 1 | | |
| 2 | Missing flag | 375 | 0.227 | 0.419 | 0 | 1 | | |
| Q8e: Barriers: Lack of data to run AI on | | | | | | | 278.18 | 0.000 |
| 0 | No | 978 | 0.593 | 0.491 | 0 | 1 | | |
| 1 | Yes | 296 | 0.180 | 0.384 | 0 | 1 | | |
| 2 | Missing flag | 375 | 0.227 | 0.419 | 0 | 1 | | |
| Q8f: Barriers: Other | | | | | | | 288.12 | 0.000 |
| 0 | No | 1202 | 0.729 | 0.445 | 0 | 1 | | |
| 1 | Yes | 72 | 0.044 | 0.204 | 0 | 1 | | |
| 2 | Missing flag | 375 | 0.227 | 0.419 | 0 | 1 | | |
| Q10: For your research, do you collaborate with scientists at firms that develop AI, such as Google, Microsoft, Tencent, Meta, IBM, Amazon, OpenAI, Baidu, or any other firm? | | | | | | | 48.41 | 0.000 |
| 0 | No | 1412 | 0.856 | 0.351 | 0 | 1 | | |
| 1 | Yes | 231 | 0.140 | 0.347 | 0 | 1 | | |
| 2 | Missing flag | 6 | 0.004 | 0.060 | 0 | 1 | | |
| Q13: How important do you think it is for researchers using AI in science to collaborate with scientists at these firms? | | | | | | | 53.12 | 0.000 |
| 1 | Very unimportant | 119 | 0.072 | 0.259 | 0 | 1 | | |
| 2 | Somewhat unimportant | 167 | 0.101 | 0.302 | 0 | 1 | | |
| 3 | Neither important nor unimportant | 362 | 0.220 | 0.414 | 0 | 1 | | |
| 4 | Somewhat important | 633 | 0.384 | 0.486 | 0 | 1 | | |
| 5 | Very important | 353 | 0.214 | 0.410 | 0 | 1 | | |
| 6 | Missing flag | 15 | 0.009 | 0.095 | 0 | 1 | | |
| Q29: Gender | | | | | | | 18.95 | 0.015 |
| 0 | Male | 1175 | 0.713 | 0.453 | 0 | 1 | | |
| 1 | Female | 393 | 0.238 | 0.426 | 0 | 1 | | |
| 2 | Missing flag | 81 | 0.049 | 0.216 | 0 | 1 | | |
| Q28r: Region | | | | | | | 28.74 | 0.026 |
| 1 | Asia | 454 | 0.275 | 0.447 | 0 | 1 | | |
| 2 | Europe | 533 | 0.323 | 0.468 | 0 | 1 | | |
| 3 | North America | 318 | 0.193 | 0.395 | 0 | 1 | | |
| 4 | Other | 219 | 0.133 | 0.339 | 0 | 1 | | |
| 5 | Missing flag | 125 | 0.076 | 0.265 | 0 | 1 | | |
| Q30: Career Stage | | | | | | | 72.13 | 0.000 |
| 1 | Early-career | 407 | 0.247 | 0.431 | 0 | 1 | | |
| 2 | Mid-career | 520 | 0.315 | 0.465 | 0 | 1 | | |
| 3 | Advanced-career | 643 | 0.390 | 0.488 | 0 | 1 | | |
| 4 | Missing flag | 79 | 0.048 | 0.214 | 0 | 1 | | |
| Q31: Work organization | | | | | | | 64.73 | 0.000 |
| 1 | Government | 102 | 0.062 | 0.241 | 0 | 1 | | |
| 2 | For-profit company | 64 | 0.039 | 0.193 | 0 | 1 | | |
| 3 | Non-profit company | 38 | 0.023 | 0.150 | 0 | 1 | | |
| 4 | University | 1057 | 0.641 | 0.480 | 0 | 1 | | |
| 5 | Hospital | 96 | 0.058 | 0.234 | 0 | 1 | | |
| 6 | Medical research institute | 128 | 0.078 | 0.268 | 0 | 1 | | |
| 7 | Government advisory group | 11 | 0.007 | 0.081 | 0 | 1 | | |
| 8 | Research funder | 33 | 0.020 | 0.140 | 0 | 1 | | |
| 9 | Other | 68 | 0.041 | 0.199 | 0 | 1 | | |
| 10 | Prefer not to say | 39 | 0.024 | 0.152 | 0 | 1 | | |
| 11 | Missing flag | 13 | 0.008 | 0.088 | 0 | 1 | | |
| Q32: Research field | | | | | | | 101.94 | 0.001 |
| 1 | Agriculture, veterinary or food science | 43 | 0.026 | 0.159 | 0 | 1 | | |
| 2 | Biological sciences | 126 | 0.076 | 0.266 | 0 | 1 | | |
| 3 | Biomedical, clinical, or health-related | 304 | 0.184 | 0.388 | 0 | 1 | | |
| 4 | Chemical sciences | 119 | 0.072 | 0.259 | 0 | 1 | | |
| 5 | Earth sciences | 58 | 0.035 | 0.184 | 0 | 1 | | |
| 6 | Economics | 22 | 0.013 | 0.115 | 0 | 1 | | |
| 7 | Engineering | 184 | 0.112 | 0.315 | 0 | 1 | | |
| 8 | Environmental sciences and Ecology | 120 | 0.073 | 0.260 | 0 | 1 | | |
| 9 | Computing or Information sciences | 191 | 0.116 | 0.320 | 0 | 1 | | |
| 10 | Mathematics | 43 | 0.026 | 0.159 | 0 | 1 | | |
| 11 | Physical sciences | 180 | 0.109 | 0.312 | 0 | 1 | | |
| 12 | Psychology | 45 | 0.027 | 0.163 | 0 | 1 | | |
| 13 | Social sciences | 93 | 0.056 | 0.231 | 0 | 1 | | |
| 14 | Humanities | 40 | 0.024 | 0.154 | 0 | 1 | | |
| 15 | Other | 65 | 0.039 | 0.195 | 0 | 1 | | |
| 16 | Missing flag | 16 | 0.010 | 0.098 | 0 | 1 | | |

The data revealed a strong correlation between the variables examined and the use of AI tools amongst researchers (see the last two columns in Table 1). In general, these results not only indicate disparities in adoption rates and perceived barriers within the scientific community, but also highlight promising collaborations between academia and industry in AI research and development.

### *4.2 Regression Analysis for Frequency of Use of Generative AI at Work*

The regression model presented in Table 2 was evaluated using several metrics. Based on data from 1,649 observations, the model showed a statistically significant overall fit, as indicated by an F-statistic of 15.39 with a probability of less than 0.001. In addition, the R-squared value of 0.248 suggests that approximately 24.8% of the variance in the dependent variable, log(Q21), is explained by the independent variables included in the model.

Table 2. Effect of variables on the frequency of AI tool use. Ordinary Least Squares (OLS) log-lin regression analysis, with robust standard errors estimated using the HC3 method, for endogenous log(Q21): "How often do you use these tools (e.g., ChatGPT, GPT-4, PALM, MidJourney, LLaMA, or any products built using these tools) at work?"

| | Coef. | Std. err. | t | P>t | Beta | Sig. | Reference category |
|---|---|---|---|---|---|---|---|
| Q1: Does your research work directly involve studying or developing AI? | | | | | | | No |
| 1. Yes | 0.121 | 0.034 | 3.57 | 0.000 | 0.110 | *** | |
| Q7: Do you feel that there are barriers preventing you, or your research team, from developing or using AI as much as you would like? | | | | | | | No |
| Yes | 0.114 | 0.052 | 2.19 | 0.028 | 0.102 | ** | |
| Missing flag | -0.340 | 0.036 | -9.54 | 0.000 | -0.259 | *** | |
| Q8a: Barriers: Lack of skills or skilled researchers | | | | | | | No |
| Yes | -0.023 | 0.043 | -0.54 | 0.590 | -0.017 | | |
| Q8b: Barriers: Lack of training resources | | | | | | | No |
| Yes | -0.084 | 0.044 | -1.92 | 0.055 | -0.058 | * | |
| Q8c: Barriers: Lack of computing resources | | | | | | | No |
| Yes | 0.030 | 0.043 | 0.71 | 0.481 | 0.021 | | |
| Q8d: Barriers: Lack of funding | | | | | | | No |
| Yes | -0.011 | 0.041 | -0.27 | 0.789 | -0.008 | | |
| Q8e: Barriers: Lack of data to run AI on | | | | | | | No |
| Yes | -0.024 | 0.041 | -0.60 | 0.549 | -0.017 | | |
| Q8f: Barriers: Other | | | | | | | No |
| Yes | 0.164 | 0.066 | 2.48 | 0.013 | 0.061 | ** | |
| Q10: For your research, do you collaborate with scientists at firms that develop AI, such as Google, Microsoft, Tencent, Meta, IBM, Amazon, OpenAI, Baidu, or any other firm? | | | | | | | No |
| 1. Yes | 0.112 | 0.036 | 3.10 | 0.002 | 0.070 | *** | |
| 2. Missing flag | 0.141 | 0.211 | 0.67 | 0.504 | 0.015 | | |
| Q13: How important do you think it is for researchers using AI in science to collaborate with scientists at these firms? | | | | | | | Very important |
| 1. Very unimportant | -0.099 | 0.057 | -1.75 | 0.080 | -0.047 | * | |
| 2. Somewhat unimportant | -0.084 | 0.048 | -1.76 | 0.078 | -0.046 | * | |
| 3. Neither important nor unimportant | -0.060 | 0.039 | -1.55 | 0.121 | -0.045 | | |
| 4. Somewhat important | -0.033 | 0.034 | -0.97 | 0.330 | -0.029 | | |
| 6. Missing flag | -0.172 | 0.151 | -1.13 | 0.257 | -0.030 | | |
| Q29: Gender | | | | | | | Male |
| 1. Female | -0.067 | 0.029 | -2.28 | 0.022 | -0.052 | ** | |
| 2. Missing flag | -0.076 | 0.067 | -1.14 | 0.254 | -0.030 | | |
| Q28r: Region | | | | | | | North America |
| 1. Asia | 0.007 | 0.037 | 0.19 | 0.849 | 0.006 | | |
| 2. Europe | -0.040 | 0.035 | -1.14 | 0.256 | -0.034 | | |
| 4. Other | 0.000 | 0.044 | 0.00 | 0.999 | 0.000 | | |
| 5. Missing flag | -0.041 | 0.052 | -0.79 | 0.429 | -0.020 | | |
| Q30: Career Stage | | | | | | | Early-career |
| 2. Mid-career | -0.117 | 0.033 | -3.50 | 0.000 | -0.098 | *** | |
| 3. Advanced-career | -0.191 | 0.033 | -5.79 | 0.000 | -0.170 | *** | |
| 4. Missing flag | -0.139 | 0.075 | -1.86 | 0.063 | -0.054 | * | |
| Q31: Working organization | | | | | | | Government |
| 2. For-profit company | 0.189 | 0.077 | 2.46 | 0.014 | 0.066 | ** | |
| 3. Non-profit company | 0.036 | 0.095 | 0.38 | 0.702 | 0.010 | | |
| 4. University | 0.134 | 0.051 | 2.64 | 0.008 | 0.117 | *** | |
| 5. Hospital | 0.153 | 0.077 | 1.99 | 0.047 | 0.065 | ** | |
| 6. Medical research institute | 0.157 | 0.067 | 2.33 | 0.020 | 0.076 | ** | |
| 7. Government advisory group | 0.447 | 0.119 | 3.77 | 0.000 | 0.066 | *** | |
| 8. Research funder | 0.063 | 0.113 | 0.56 | 0.576 | 0.016 | | |
| 9. Other | 0.040 | 0.082 | 0.48 | 0.628 | 0.014 | | |
| 10. Prefer not to say | 0.023 | 0.096 | 0.24 | 0.813 | 0.006 | | |
| 11. Missing flag | 0.143 | 0.167 | 0.86 | 0.391 | 0.023 | | |
| Q32: Research field | | | | | | | Computer or Information sciences |
| 1. Agriculture, veterinary or food science | -0.120 | 0.081 | -1.48 | 0.138 | -0.035 | | |
| 2. Biological sciences | -0.061 | 0.059 | -1.04 | 0.297 | -0.030 | | |
| 3. Biomedical, clinical, or health-related sciences | -0.095 | 0.051 | -1.87 | 0.061 | -0.067 | * | |
| 4. Chemical sciences | -0.080 | 0.061 | -1.31 | 0.191 | -0.038 | | |
| 5. Earth sciences | -0.103 | 0.083 | -1.24 | 0.217 | -0.034 | | |
| 6. Economics | -0.028 | 0.125 | -0.23 | 0.821 | -0.006 | | |
| 7. Engineering | -0.047 | 0.054 | -0.87 | 0.383 | -0.027 | | |
| 8. Environmental sciences and Ecology | -0.180 | 0.060 | -2.98 | 0.003 | -0.085 | *** | |
| 10. Mathematics | -0.157 | 0.082 | -1.92 | 0.055 | -0.045 | * | |
| 11. Physical sciences | -0.066 | 0.056 | -1.20 | 0.232 | -0.038 | | |
| 12. Psychology | 0.101 | 0.078 | 1.31 | 0.192 | 0.030 | | |
| 13. Social sciences | -0.009 | 0.065 | -0.14 | 0.889 | -0.004 | | |
| 14. Humanities | 0.030 | 0.086 | 0.35 | 0.725 | 0.008 | | |
| 15. Other | -0.108 | 0.073 | -1.49 | 0.135 | -0.038 | | |
| 16. Missing flag | -0.051 | 0.147 | -0.35 | 0.729 | -0.009 | | |
| _cons | 0.796 | 0.083 | 9.59 | 0.000 | . | *** | |
| Number of obs = | 1649 | | | | | | |
| F(50, 1598) = | 15.39 | | | | | | |
| Prob > F = | 0.000 | | | | | | |
| R-squared = | 0.2479 | | | | | | |
| Root MSE = | 0.48456 | | | | | | |

*Note: Statistically significant at 1% (***), 5% (**) or 10% (*)*

The model revealed several critical factors influencing researchers' use of AI tools. As expected, researchers directly involved in AI research and development showed a significant propensity to frequently use AI tools (coef: 0.121, p-value < 0.001), corresponding to a 12% increase. Interestingly, researchers who perceived general barriers to AI adoption, as well as those who encountered other barriers, were also more likely to use AI tools frequently (coef: 0.114 and 0.164, p-values: 0.028 and 0.013, respectively), an 11% and 16% increase over those who reported no barriers (Figure 2). This suggests that perceived barriers may serve as a motivation for researchers to integrate AI tools into their workflow, possibly driven by the recognition of potential benefits and a proactive stance in addressing challenges.

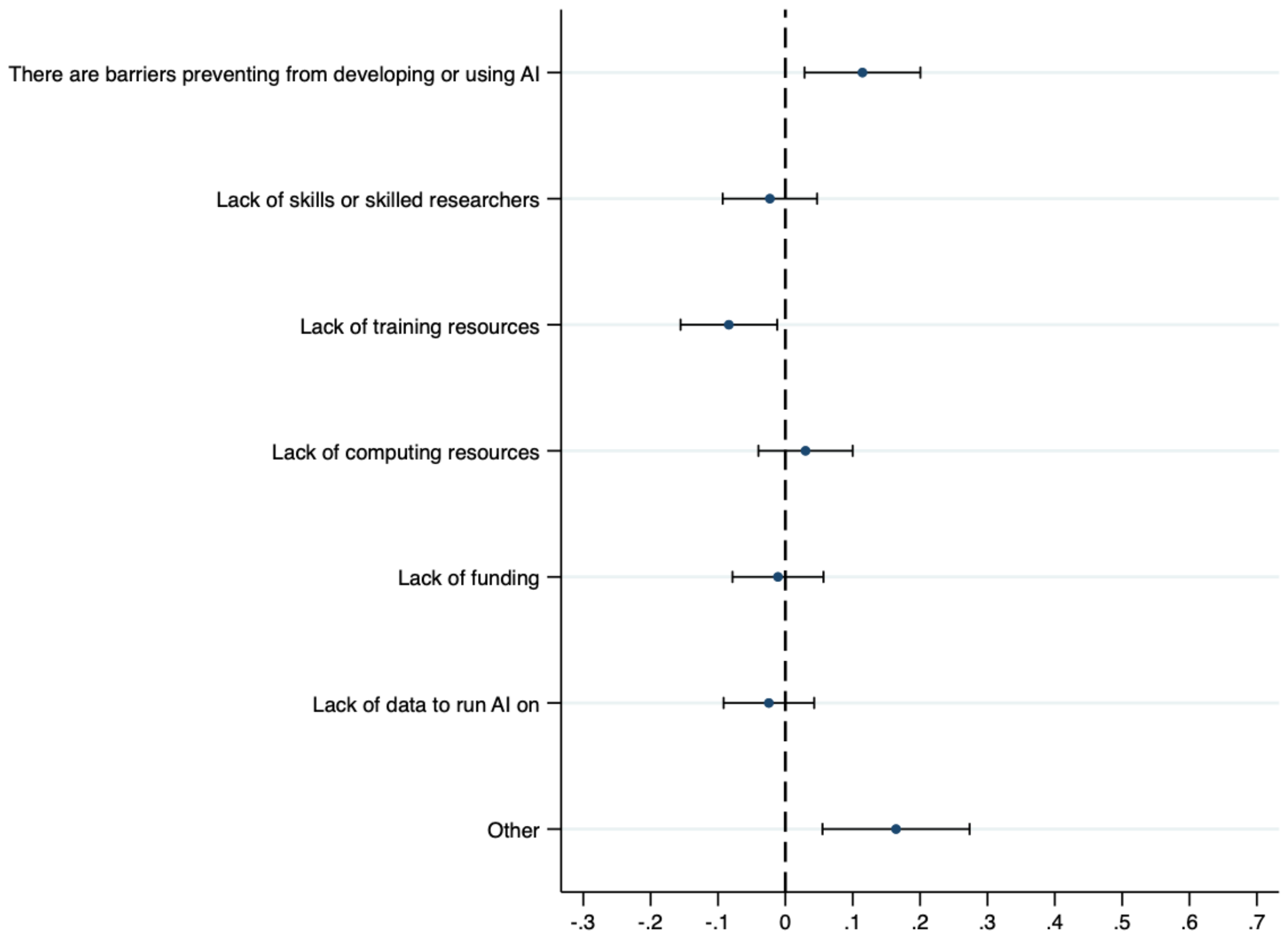

Figure 2. Estimated coefficients and confidence intervals for variables related to researchers' perceived barriers to developing or using AI (Reference group: No barriers)

Conversely, a negative association was observed with a lack of training resources (coef: -0.084, p-value: 0.055), indicating an 8% decrease in the frequent use of AI tools amongst researchers who identified this barrier (Figure 2). This is consistent with the expected negative impact of specific barriers. Overall, these findings highlight the complex relationship between perceived barriers and technology adoption, where challenges can either hinder or drive researchers' engagement with emerging technologies such as AI.

The model also identified a positive correlation between collaborating with AI development companies and frequent use of AI tools (coef: 0.112, p-value: 0.002), implying an 11% increase in researchers engaging in such collaborations. Conversely, researchers who perceived these collaborations as very or somewhat unimportant had lower rates of AI tool use (negative coef: -0.099 and -0.084, p-values < 0.08), corresponding to decreases of 10% and 8%, respectively.

Furthermore, a clear trend emerged regarding career stage, with early-career researchers showing higher rates of frequent AI tool use compared to mid-career and advanced-career counterparts (coef: -0.117 and -0.191, respectively; p-values < 0.001), indicating 12% and 19% more use, respectively (Figure 3). This finding could be attributed to several factors, including greater familiarity with the technology, more recent training incorporating AI advances, pressure to publish, openness to new methodologies, and access to research funding.

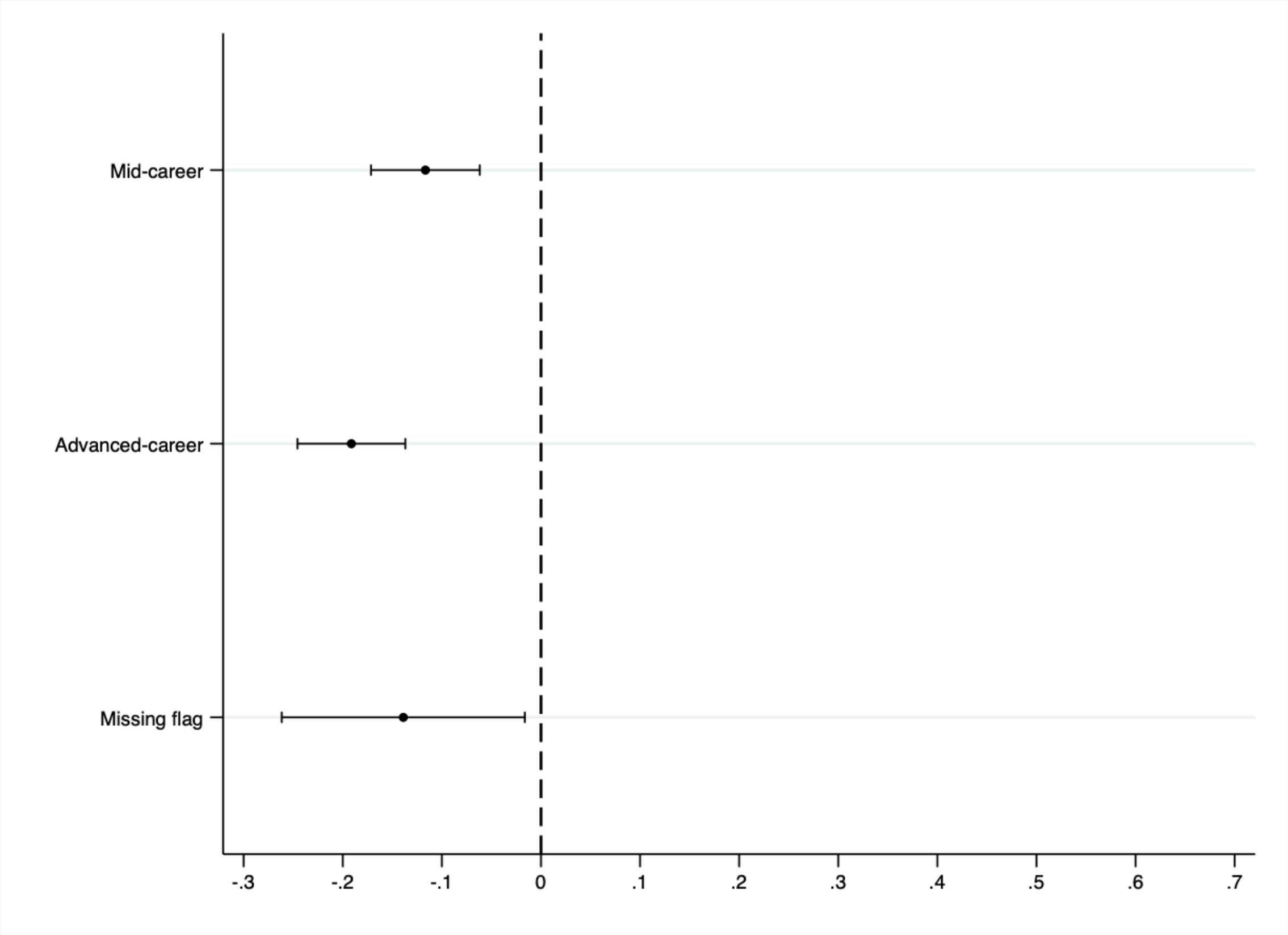

Figure 3. Estimated coefficients and confidence intervals for variables related to researcher career stage (Reference group: Early-career)

In terms of gender, a statistically significant negative association was observed for female researchers (coef: -0.067, p-value: 0.022), indicating a 7% decrease in the frequent use of AI tools compared to male researchers. However, the effect size was relatively small and warrants further investigation into the underlying reasons.

Analysis by type of research organization revealed higher rates of frequent AI tool use amongst researchers at for-profit companies, medical research institutes, hospitals, and universities compared to government positions, with increases ranging from 19% to 13% (Figure 4). This trend may be due to a focus on innovation and available resources in these sectors.

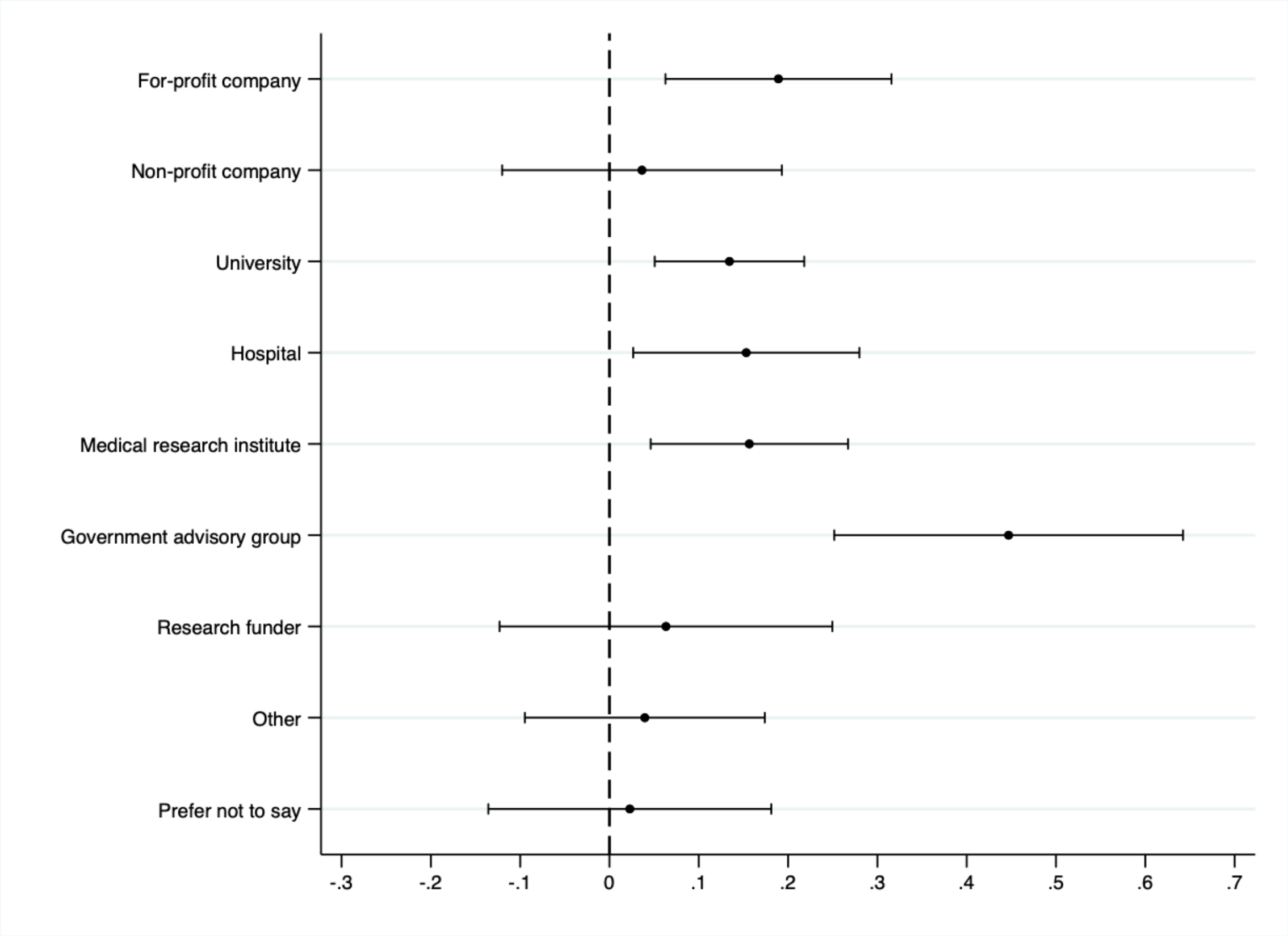

Figure 4. Estimated coefficients and confidence intervals for variables related to researcher working organization (Reference group: Government)

In addition, significant associations were found between the field of study and the frequency of use of AI tools. In particular, researchers in environmental sciences and ecology, mathematics, and biomedical, clinical, and health sciences showed decreases of 18%, 16%, and 9%, respectively, compared to computer and information sciences (Figure 5).

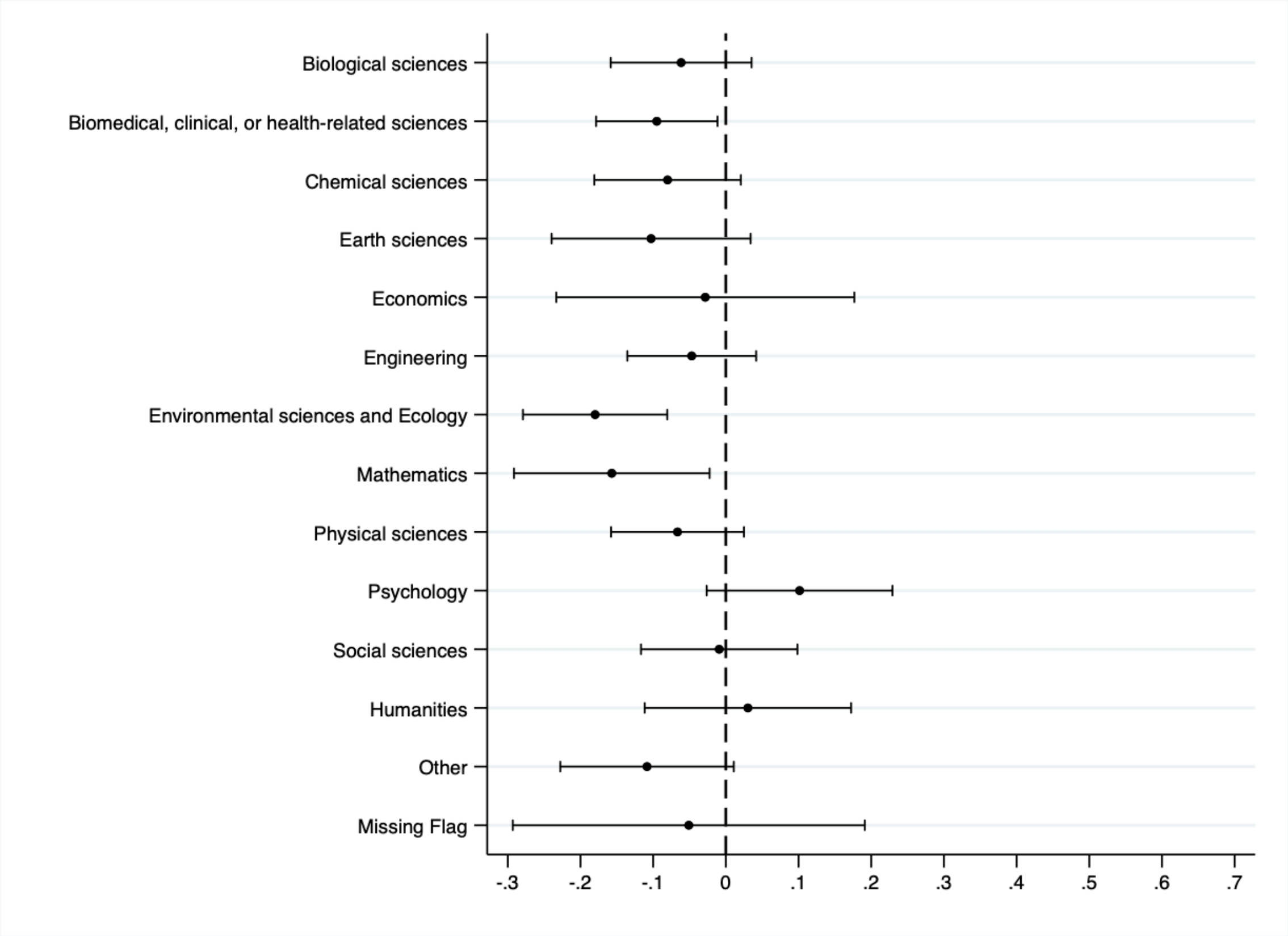

Figure 5. Estimated coefficients and confidence intervals for variables related to the field of research (Reference group: Computer or Information sciences)

The geographic location did not show significant differences in the frequency of use of AI tools amongst researchers in North America, Asia, or Europe. However, it is important to consider broader geographic diversity for a more comprehensive analysis.

In conclusion, while certain factors such as career stage, collaboration, gender, workplace type, and perceived barriers influence researchers' engagement with AI tools, further research is warranted to understand the nuanced dynamics that shape technology adoption in research settings.

## 5. Discussion

The Nature survey on AI and science involved more than 1,600 researchers worldwide and explored their views on the increasing use of artificial intelligence tools in research. AI has transformed research by enabling faster data processing, speeding up computations, saving time and money, and facilitating progress in answering complex questions that were previously challenging. However, concerns have been raised about potential negative impacts, such as increased reliance on pattern recognition without understanding, entrenching bias or discrimination in data, making it easier to perpetrate fraud, and challenging existing standards of proof and truth in science (Van Noorden and Perkel, 2023). The survey highlighted the need for researchers to carefully navigate the benefits and risks of AI in science to ensure its safe integration into research practices and society (Nature, 2023a).

However, the results of the survey may be skewed by a disproportionate representation of scientists already interested in AI, which could introduce bias by overemphasizing the viewpoints of AI researchers. To address this concern, the current study employs a rigorous methodology that goes beyond basic descriptive data analysis. It uses an inferential technique, regression analysis, to isolate effects. This approach allows researchers to disentangle the independent effects of various factors such as gender, career stage, institutional affiliation, and perceived barriers to the adoption of generative AI. In addition, the study accounts for potential confounding variables such as direct involvement in AI research, collaboration with AI-focused companies, geographic location, and scientific discipline. Through this comprehensive strategy, the study provides a more nuanced understanding of the factors influencing the adoption of generative AI in the scientific community at large overcoming the potential biases inherent in AI-centric circles.

Based on effect size, researchers actively engaged in AI research or development show a 12% increase in the use of AI tools compared to their peers. Collaborators with AI companies show an 11% increase, while those who consider such collaborations very or somewhat unimportant show a 10% and 8% decrease in AI tool usage, respectively, compared to those who consider collaboration very important. In addition, researchers who perceived barriers to AI implementation experience an 11% increase in tool usage compared to those without perceived barriers. Conversely, those who cite inadequate training resources experience an 8% decrease, while those who face other types of barriers show a 16% increase in AI tool use.

Female researchers show a 7% reduction in AI tool use relative to their male counterparts. Advanced-career researchers show a significant 19% decrease, while mid-career researchers show a 12% decrease compared to early-career researchers. In contrast to researchers in government roles, those associated with government advisory groups are 45% more likely to use AI tools frequently. In addition, researchers at for-profit companies show a 19% increase, while those at medical research institutions and hospitals show a 16% and 15% increase, respectively. University researchers also show a 13% increase in the use of AI tools. Finally, researchers in environmental sciences and ecology show an 18% decrease in the use of AI tools compared to researchers in computer or information sciences. Mathematicians show a 16% decrease, and researchers in biomedical, clinical, or health sciences see a 9% decrease in AI tool usage.

Some reflections on these findings include the following. Interestingly, we found that researchers who perceive barriers to AI adoption (general or specific) are more likely to use AI tools. This may be due to their motivation to overcome the challenges and realize the potential benefits of AI (overcoming challenges effect). As an exception to this trend, researchers lacking training resources were less likely to use AI tools frequently, highlighting the importance of training for adoption. These results show the complexity of human behaviour when faced with barriers in the workplace (see Hangl et al., 2023; Li et al., 2024).

Interestingly, we also found that early-career researchers were significantly more likely to use AI tools than mid-career and senior researchers (early career advantage). This could be due to several factors: comfort with new technologies, recent training in advances such as AI, pressure to publish and need for efficient workflows, openness to exploring new approaches, and limited access to established labs, which makes AI tools (especially open-source or cloud-based) attractive.

Researchers working in for-profit companies, medical institutions, hospitals, and universities are more likely to use AI tools than those working in government (Organizational Influence). This may be due to a focus on innovation and potential funding for AI research in these sectors (see Berman et al., 2024). The strongest association is with government advisory groups, likely due to their position at the forefront of emerging technologies and access to significant resources.

We found a small, statistically significant negative association between being female and frequent use of AI tools (gender gap). However, more research is needed to understand this. Researchers in environmental sciences, mathematics, and biomedical sciences use AI tools less frequently than those in computer or information sciences (field effect). In addition, no significant differences in the frequency of AI tool use were found between North America, Asia, and Europe (no geographic location effect). However, the study only looked at a limited number of regions, so a more comprehensive analysis may be needed.

The findings presented in this paper have several policy implications that could shape the future integration of generative AI technologies in research environments. The results highlight critical inequalities and barriers that influence the adoption and use of AI tools among researchers, suggesting targeted policy interventions to promote equitable and effective AI integration.

The observed 7% lower use of AI tools among female researchers compared to their male counterparts highlights the need for gender-sensitive policies in AI training and resource allocation. Policymakers should consider implementing initiatives that specifically support female researchers, such as targeted AI training programs, mentorship opportunities, and the establishment of supportive networks that encourage the adoption of AI tools. In addition, addressing systemic biases within the research community and promoting an inclusive culture can help close this gap.

The significant 19% decrease in the use of AI tools among advanced career researchers suggests a potential resistance or lack of need for AI integration among more experienced researchers. Policy measures could include the provision of training and retraining programs tailored to the needs of senior researchers. These programs should highlight the benefits and practical applications of generative AI tools in enhancing research productivity and innovation, thereby encouraging experienced researchers to embrace new technologies.

The finding that researchers who face barriers to AI adoption are 11% more likely to use AI tools suggests that overcoming initial barriers can lead to increased engagement with AI technologies. Policies should aim to identify and mitigate these barriers by providing comprehensive support structures, such as user-friendly AI platforms, accessible training resources, and dedicated technical support. Conversely, the 8% decrease in AI tool usage due to insufficient training resources highlights the critical need for robust and widely available educational materials and training programs to ensure that all researchers can effectively use AI tools.

The different levels of AI tool usage across different types of workplaces - higher in government advisory groups (45%), for-profit companies (19%), and medical research institutions (16%) - indicate that institutional context plays a critical role in AI adoption. Policies should encourage cross-sector collaboration and knowledge sharing to disseminate best practices and successful AI integration strategies. For example, fostering partnerships between academic institutions and industry can help bridge gaps in AI applications and provide researchers with diverse perspectives and resources.

The lower frequency of use of AI tools among researchers in government roles compared to government advisory groups suggests a need for targeted interventions within government research organizations. Policies should focus on improving the infrastructure and support for AI adoption in these environments, such as providing dedicated funding for AI initiatives, encouraging interdepartmental collaboration, and establishing clear guidelines and frameworks for AI use in public sector research.

The findings call for a comprehensive and multi-faceted policy approach to promote the widespread and equitable adoption of generative AI tools in research. This includes investing in education and training, addressing gender and career stage inequalities, supporting diverse institutional contexts, and continuously assessing and addressing perceived barriers to AI integration. By adopting these measures, policymakers can ensure that all researchers, regardless of gender, career stage, or workplace, can fully exploit the transformative potential of generative AI technologies in their work.

**References**

Agathokleous, E., Rillig, M. C., Peñuelas, J., & Yu, Z. (2024). One hundred important questions facing plant science derived using a large language model. Trends in Plant Science, 29(2). https://doi.org/10.1016/j.tplants.2023.06.008

Alkaissi, H., & McFarlane, S. I. (2023). Artificial hallucinations in ChatGPT: implications in scientific writing. Cureus, 15(2), e35179. https://doi.org/10.7759/cureus.35179

Altmäe, S., Sola-Leyva, A., & Salumets, A. (2023). Artificial intelligence in scientific writing: a friend or a foe?. Reproductive BioMedicine Online, 47(1), 3-9. https://doi.org/10.1016/j.rbmo.2023.04.009

Berman, A., de Fine Licht, K., & Carlsson, V. (2024). Trustworthy AI in the public sector: An empirical analysis of a Swedish labor market decision-support system. Technology in Society, 76, 102471. https://doi.org/10.1016/j.techsoc.2024.102471

Bin-Nashwan, S. A., Sadallah, M., & Bouteraa, M. (2023). Use of ChatGPT in academia: Academic integrity hangs in the balance. Technology in Society, 75, 102370. https://doi.org/10.1016/j.techsoc.2023.102370

Biswas, S. (2023). ChatGPT and the future of medical writing. Radiology, 307(2), e223312. https://doi.org/10.1148/radiol.223312

Briganti, G., & Le Moine, O. (2020). Artificial intelligence in medicine: today and tomorrow. Frontiers in Medicine, 7, 509744. https://doi.org/10.3389/fmed.2020.00027

Chen, T. J. (2023). ChatGPT and other artificial intelligence applications speed up scientific writing. Journal of the Chinese Medical Association, 86(4), 351-353. https://doi.org/10.1097/JCMA.0000000000000900

Davidson, R., & MacKinnon, J.G. (1993). *Estimation and Inference in Econometrics*. New York: Oxford University Press.

Davidson, R., & MacKinnon, J.G. (2004). *Econometric Theory and Methods*. New York: Oxford University Press.

De Angelis, L., Baglivo, F., Arzilli, G., Privitera, G. P., Ferragina, P., Tozzi, A. E., & Rizzo, C. (2023). ChatGPT and the rise of large language models: the new AI-driven infodemic threat in public health. Frontiers in Public Health, 11, 1166120. https://doi.org/10.3389/fpubh.2023.1166120

ERC (2023). Foresight: Use and impact of Artificial Intelligence in the scientific process. https://erc.europa.eu/sites/default/files/2023-12/AI_in_science.pdf

Floridi, L., & Chiriatti, M. (2020). GPT-3: Its nature, scope, limits, and consequences. Minds and Machines, 30, 681-694. https://doi.org/10.1007/s11023-020-09548-1

Gao, A., Murphy, R. R., Chen, W., Dagnino, G., Fischer, P., Gutierrez, M. G., ... & Yang, G. Z. (2021). Progress in robotics for combating infectious diseases. Science Robotics, 6(52), eabf1462. https://doi.org/10.1126/scirobotics.abf1462

Gao, C. A., Howard, F. M., Markov, N. S., Dyer, E. C., Ramesh, S., Luo, Y., & Pearson, A. T. (2023). Comparing scientific abstracts generated by ChatGPT to real abstracts with detectors and blinded human reviewers. NPJ Digital Medicine, 6(1), 75. https://doi.org/10.1038/s41746-023-00819-6

Glickman, M., & Zhang, Y. (2024). AI and Generative AI for Research Discovery and Summarization. Harvard Data Science Review, 6(2). https://doi.org/10.1162/99608f92.7f9220ff

Grace, K., Stewart, H., Sandkühler, J. F., Thomas, S., Weinstein-Raun, B., & Brauner, J. (2024). Thousands of AI authors on the future of AI. arXiv preprint arXiv:2401.02843.

Hangl, J., Krause, S., & Behrens, V. J. (2023). Drivers, barriers and social considerations for AI adoption in SCM. Technology in Society, 74, 102299. https://doi.org/10.1016/j.techsoc.2023.102299

Holler, J., & Levinson, S. C. (2019). Multimodal language processing in human communication. Trends in Cognitive Sciences, 23(8), 639-652. https://doi.org/10.1016/j.tics.2019.05.006

Huang, J., & Tan, M. (2023). The role of ChatGPT in scientific communication: writing better scientific review articles. American Journal of Cancer Research, 13(4), 1148–1154. https://www.ncbi.nlm.nih.gov/pmc/articles/PMC10164801

Hutson, M. (2022). Could AI help you to write your next paper?. Nature, 611(7934), 192-193. https://www.nature.com/articles/d41586-022-03479-w

Kung, T. H., Cheatham, M., Medenilla, A., Sillos, C., De Leon, L., Elepaño, C., ... & Tseng, V. (2023). Performance of ChatGPT on USMLE: potential for AI-assisted medical education using large language models. PLoS Digital Health, 2(2), e0000198. https://doi.org/10.1371/journal.pdig.0000198

Lee, J. Y. (2023). Can an artificial intelligence chatbot be the author of a scholarly article?. Journal of Educational Evaluation for Health Professions, 20, 6. https://doi.org/10.3352/jeehp.2023.20.6

Li, Y., Song, Y., Sun, Y., & Zeng, M. (2024). When do employees learn from artificial intelligence? The moderating effects of perceived enjoyment and task-related complexity. Technology in Society, 102518. https://doi.org/10.1016/j.techsoc.2024.102518

Marquez, R., Barrios, N., Vera, R. E., Mendez, M. E., Tolosa, L., Zambrano, F., & Li, Y. (2023). A perspective on the synergistic potential of artificial intelligence and product-based learning strategies in biobased materials education. Education for Chemical Engineers, 44, 164-180. https://doi.org/10.1016/j.ece.2023.05.005

Nature (2023a). AI will transform science — now researchers must tame it. https://www.nature.com/articles/d41586-023-02988-6

Nature (2023b). Is AI leading to a reproducibility crisis in science? https://www.nature.com/articles/d41586-023-03817-6

Nature (2023c). Science and the new age of AI. https://www.nature.com/immersive/d41586-023-03017-2/index.html.

Neumeister, L. (2023, June 9). Lawyers blame ChatGPT for tricking them into citing bogus case law. Associated Press News. https://apnews.com/article/artificial-intelligence-chatgpt-courts-e15023d7e6fdf4f099aa122437dbb59b

Norris, C. (2023). Large language models like ChatGPT in ABME: author guidelines. Annals of Biomedical Engineering, 51(6), 1121-1122. https://doi.org/10.1007/s10439-023-03212-2

Rahimi, F., & Abadi, A. T. B. (2023). ChatGPT and publication ethics. Archives of Medical Research, 54(3), 272-274. https://doi.org/10.1016/j.arcmed.2023.03.004

Salvagno, M., Taccone, F. S., & Gerli, A. G. (2023). Can artificial intelligence help for scientific writing?. Critical Care, 27(1), 75. https://doi.org/10.1186/s13054-023-04380-2

Thirunavukarasu, A. J., Ting, D. S. J., Elangovan, K., Gutierrez, L., Tan, T. F., & Ting, D. S. W. (2023). Large language models in medicine. Nature Medicine, 29(8), 1930-1940. https://doi.org/10.1038/s41591-023-02448-8

Trenfield, S. J., Awad, A., McCoubrey, L. E., Elbadawi, M., Goyanes, A., Gaisford, S., & Basit, A. W. (2022). Advancing pharmacy and healthcare with virtual digital technologies. Advanced Drug Delivery Reviews, 182, 114098. https://doi.org/10.1016/j.addr.2021.114098

Van Noorden, R., & Perkel, J. M. (2023). AI and science: what 1,600 researchers think. Nature, 621(7980), 672-675. https://doi.org/10.1038/d41586-023-02980-0

Wang, H., Fu, T., Du, Y., Gao, W., Huang, K., Liu, Z., ... & Zitnik, M. (2023). Scientific discovery in the age of artificial intelligence. Nature, 620(7972), 47-60. https://doi.org/10.1038/s41586-023-06221-2

Xu, Y., Liu, X., Cao, X., Huang, C., Liu, E., Qian, S., ... & Zhang, J. (2021). Artificial intelligence: A powerful paradigm for scientific research. The Innovation, 2(4), 100179. https://doi.org/10.1016/j.xinn.2021.100179